\documentclass[longauth]{aa} 
\usepackage{natbib}
\bibpunct{(}{)}{;}{a}{}{,}
\usepackage{graphicx}
\usepackage{multirow}
\usepackage{mathabx}
\usepackage{upgreek}

\usepackage{txfonts}

\usepackage{hyperref} \hypersetup{ colorlinks=true, citecolor={blue} ,linkcolor=blue, filecolor=blue, urlcolor=blue, pdftitle={Overleaf Example}}

\newcommand{\muas}{$\upmu\rm{as}$}

\newcommand{\msun}{M$_{\odot}$}
\newcommand{\msunpyr}{M$_{\odot}$.yr$^{-1}$}
\newcommand{\Lsun}{L$_\odot$}
\newcommand{\rsun}{R$_{\odot}$}

\newcommand{\ddeg}{$^{\circ}$}
\newcommand{\te}{T$_{\rm eff}$}

\newcommand{\prot}{P$_{\rm rot}$}
\newcommand{\Brgam}{Br$\gamma$}
\newcommand{\Th}{$^\text{th}$}

\begin{document} 
\nolinenumbers
      \title{The GRAVITY young stellar object survey}
   \subtitle{XIV. Investigating the magnetospheric accretion-ejection processes in S~CrA~N\thanks{GTO programs with run ID 109.22ZC.004}}
   


   \author{GRAVITY Collaboration: H. Nowacki\inst{1}
        \and    K.~Perraut\inst{1}
        \and    L.~Labadie\inst{2}
        \and    J.~Bouvier\inst{1}
        \and    C.~Dougados\inst{1}
        \and    M.~Benisty\inst{1,3}
        \and    J.~A.~Wojtczak\inst{2}
        \and    A.~Soulain\inst{1}
        \and    E.~Alecian\inst{1}
        \and    W.~Brandner\inst{5}
        \and    A.~Caratti~o~Garatti\inst{5,6,7}
        \and    R.~Garcia~Lopez\inst{5,7}
        \and    V.~Ganci\inst{2,4}
        \and    J.~S\'anchez-Berm\'udez\inst{5,8}
        \and    J.-P.~Berger\inst{1}
        \and    G.~Bourdarot\inst{9}
        \and    P.~Caselli\inst{9}
        \and    Y.~Clénet\inst{10}
        \and    R.~Davies\inst{9}
        \and    A.~Drescher\inst{9}
        \and    A.~Eckart\inst{2,4}
        \and    F.~Eisenhauer\inst{9}
        \and    M.~Fabricius\inst{9}
        \and    H.~Feuchtgruber\inst{9}
        \and    N.~M.~Förster-Schreiber\inst{9}
        \and    P.~Garcia\inst{11,12}
        \and    E.~Gendron\inst{10}
        \and    R.~Genzel\inst{9}
        \and    S.~Gillessen\inst{9}
        \and    S.~Grant\inst{9}
        \and    T.~Henning\inst{5}
        \and    L.~Jocou\inst{1}
        \and    P.~Kervella\inst{10}
        \and    N.~Kurtovic\inst{9}
        \and    S.~Lacour\inst{10}
        \and    V.~Lapeyrère\inst{10}
        \and    J.-B.~Le~Bouquin \inst{1}
        \and    D.~Lutz\inst{9}
        \and    F.~Mang\inst{9}
        \and    T.~Ott\inst{9}
        \and    T.~Paumard\inst{10}
        \and    G.~Perrin\inst{10}
        \and    S.~Rabien\inst{9}
        \and    D.~Ribeiro\inst{9}
        \and    M.~Sadun~Bordoni\inst{9}
        \and    S.~Scheithauer\inst{5}
        \and    J.~Shangguan\inst{9}
        \and    T.~Shimizu\inst{9}
        \and    S.~Spezzano\inst{9}
        \and    C.~Straubmeier\inst{2}
        \and    E.~Sturm\inst{9}
        \and    L.~Tacconi\inst{9}
        \and    E.~van~Dishoeck\inst{9}
        \and    F.~Vincent\inst{10}
        \and    F.~Widmann\inst{9}
        }

\institute{Univ. Grenoble Alpes, CNRS, IPAG, 38000 Grenoble, France \\
\email{hugo.nowacki@univ-grenoble-alpes.fr}
\and I. Physikalisches Institut, Universität zu Köln, Zülpicher Str. 77, 50937, Köln, Germany
\and Université Côte d'Azur, Observatoire de la Côte d'Azur, CNRS, Laboratoire Lagrange, France
\and Max-Planck-Institute for Radio Astronomy, Auf dem Hügel 69, 53121 Bonn, Germany
\and Max Planck Institute for Astronomy, Königstuhl 17, 69117 Heidelberg, Germany 
\and INAF-Osservatorio Astronomico di Capodimonte, Salita Moiariello 16, 80131 Napoli, Italy 
\and School of Physics, University College Dublin, Belfield, Dublin 4, Ireland
\and Instituto de Astronomia, Universidad Nacional Autonoma de Mexico, Apdo. Postal 70264, Ciudad de Mexico, 04510, Mexico
\and Max Planck Institute for Extraterrestrial Physics, Giessenbachstrasse, 85741 Garching bei München, Germany
\and LESIA, Observatoire de Paris, PSL Research University, CNRS, Sorbonne Universités, UPMC Univ. Paris 06, Univ. Paris Diderot, Sorbonne Paris Cité, France
\and  CENTRA, Centro de Astrofísica e Gravitação, Instituto Superior Técnico, Avenida Rovisco Pais 1, 1049 Lisboa, Portugal 
\and Universidade do Porto, Faculdade de Engenharia, Rua Dr. Roberto Frias, 4200-465 Porto, Portugal
}

   \date{Received June 26, 2024; accepted August 1, 2024}

  \abstract
   {The dust- and gas-rich protoplanetary disks around young stellar systems play a key role in star and planet formation. While considerable progress has recently been made in probing these disks on large scales of a few tens of astronomical units (au), the central au needs to be more investigated.
   }
   {We aim at unveiling the physical processes at play in the innermost regions of the strongly accreting T Tauri Star S CrA N by means of near-infrared interferometric observations. As recent spectropolarimetric observations suggested that S~CrA~N might undergo intense ejection processes, we focus on the accretion-ejection phenomena and on the star-disk interaction region.
   }
   {We obtained interferometric observations with VLTI/GRAVITY in the K-band during two consecutive nights in August 2022. The analysis of the continuum emission, coupled with the differential analysis across 
the \Brgam~line, allows us to constrain the morphology of the dust and gas distribution in the innermost regions of S~CrA~N and to investigate their temporal variability. These observations are compared to magnetospheric accretion/ejection models of T Tauri stars and to previous observations to describe the physical processes operating in these regions.}
   {The K-band continuum emission is well reproduced with an azimuthally-modulated dusty ring with a half-light radius of 0.24~au ($\sim$~20~$R_*$), an inclination of $\sim$30\ddeg, and a position angle of $\sim$150\ddeg. As the star alone cannot explain such a large sublimation front, we propose that magnetospheric accretion is an important dust-heating mechanism leading to this continuum emission. The \Brgam~emitting region (0.05-0.06~au; 5-7~$R_*$) is found to be more compact than the continuum, and of the order or larger than the size of the magnetospheric truncation radius. The on-sky displacements across the \Brgam~spectral channels are aligned along a position angle offset by 45\ddeg{} from the disk, and extend up to 2 $R_*$. This is in agreement with radiative transfer models combining magnetospheric accretion and disk winds. These on-sky displacements remain unchanged from one night to the other, while the line flux decreases by 13\%, suggesting a dominant contribution of wind for the origin of the \Brgam~line.
   }
   {Our observations support an origin of the \Brgam~line from a combination of (variable) accretion-ejection processes in the inner disk region. 
   }

     \keywords{Stars: variables: T Tauri --
                Stars: individual: S~CrA~N --
                Techniques: interferometry --
                Accretion, accretion disk
               }

   \maketitle

\section{Introduction}

The young solar-mass stars, the so-called Classical T Tauri Stars (CTTS), are still accreting matter from their circumstellar disks and are the site of intense magnetospheric accretion-ejection processes. Notably, within the framework of the magnetospheric accretion scenario, the stellar magnetic field truncates the inner disk close to the central star (i.e., at a few hundredths of an astronomical unit (au) away) and forces matter to be funneled onto the star along the magnetic field lines, creating accretion shocks at the stellar surface \citep{bouvier_2007,Hartmann_2016}. These events are responsible for the emission-rich spectra of CTTS, which in some cases display exceptionally intense and variable lines \citep{White2003}. Temporal variability of {photometry} is very frequently observed in CTTS on different timescales, ranging from sub-daily to monthly periods, regardless if accretion is strong (e.g., \cite{Siwak2016_RULup} for RU~Lup) or not (e.g., \cite{Rucinski2008_TWHya,Siwak2011_TWHya} for TW~Hya). The variations in shape and in intensity of emission lines can be stochastic, episodic or periodic \citep[see][for a review of accretion variabilities]{fischer_2023}. The origin of this variability is still poorly constrained and requests multi-technique or multi-wavelength campaigns to be understood. Such a simultaneous multi-technique approach has recently proven effective in providing unique constraints for the CTTS DoAr~44: while spectro-polarimetry (with ESPaDOnS and SPIRou at CFHT) allowed for kinematics studies across many (variable) spectral lines and for monitoring the magnetic field topology \citep{bouvier_2020a}, near-infrared spectro-interferometry across the \Brgam~line with GRAVITY at VLTI \citep{gravity_collaboration_first_2017} allowed to resolve the Hydrogen emitting region at a scale of a few stellar radii 
\citep{bouvier_2020b}. 
Combining the different techniques is helpful for disentangling the contribution of accretion and ejection flows to the \Brgam~line \citep{gravity_2020,gravity_2023_CITau,gravity_collaboration_gravity_2023}.

We apply this multi-technique approach to the strongly accreting T Tauri star S~CrA~North (hereafter, S~CrA~N). The strong accretors generally exhibit more intense and more variable emission lines (e.g., \cite{Venuti2014_NGC2264} for the objects of the NGC~2264 cluster). As they are often deeply embedded in an complex environment, they are difficult to observe, notably in the visible/near-infrared range. Due to their high accretion luminosity, the central objects are highly veiled. As such, they have been poorly studied, and it is still unclear whether the strong accretor status corresponds to a specific evolutionary stage or a specific geometric configuration of CTTS.

S~CrA~N is the Northern component of the S~CrA binary system, located near the most extincted region of the Corona Australis cloud at a distance of $\sim$~150~pc \citep{dobashi_2005}. The two components are observed close to each other \citep[1.3 arcsec, see][]{zhang_2023}, they are similar in mass and are coeval, with spectral type K6 \citep[][and references therein]{gahm_2018}. S~CrA has been identified as a variable T Tauri binary since \citet{joy_1945}, and as a bright member of the YY~Ori class by \citet{walker_1972}. The two components present broad redshifted absorptions in various emission lines \citep[e.g. Fe~II~$\lambda$4924,$\lambda$5018,$\lambda$5169, He~I~D$_3$, higher Balmer lines, see][]{gahm_2018,nowacki_star-disk_2023}, which is indicative of material falling onto the central objects at free-fall velocities \citep{krautter_1990}, 
thus offering a first clue of the strong accretor nature of this binary system \citep{petrov_2014}. Many studies later confirmed this status in a variety of manners: the spectral features observed in its spectrum have never been fully reproduced, but the most promising results were obtained using strong accretion models \citep[][]{gahm_2018} in combination with intense chromospheric emission models \citep{petrov_2011, dodin_2012}. Evidence has been found for a very dust- and gas-rich environment indicative of a remnant envelope at large scales \citep{sicilia-aguilar_2013} and a massive dusty disk ($\sim$102 $M_{\Earth}$ and $\sim$100 au in radius) is observed around S~CrA~N \citep{cazzoletti_2019,zhang_2023}. Altogether with a flat spectral energy distribution \citep[e.g.][]{Sullivan_2019}, these elements indicate a very young age {as discussed in \cite{nowacki_star-disk_2023} (henceforth N23), who led an observing campaign in optical spectropolarimetry with CFHT/ESPaDOnS to derive the star's fundamental parameters (Table~\ref{tab:param}) and magnetic field topology.} This system was also studied by near-infrared interferometry and the previous observations showed the presence of a bright sublimation front at the inner rim of the disk \citep[0.13--0.15 au, when considering a distance of 152.4 pc;][]{vural_2012, gravity_collaboration_wind_2017}.

In this paper, we report on a study of this inner rim and its inward regions, thanks to a richer and higher-quality GRAVITY data set, that show significant differential signals across the \Brgam~line during two consecutive nights.
Section \ref{Sect:Observations} details the GRAVITY observations and the data reduction we applied. Section \ref{Sec:Results} presents our results on the K-band continuum and Br$\gamma$ line emitting regions that are discussed in Section \ref{Sec:Discussion}. Section \ref{Sec:Conclusions} concludes on the interpretation of these results in the context of S~CrA~N and strong accretors in general.

\begin{table}
    \caption{Parameters of S~CrA~N relevant for this work. All reference values come from N23.}
    \label{tab:param}
    \centering
    \begin{tabular}{l c}
    \hline
    \hline
    Parameters & Reference value \\
    \hline
    $d$ [pc] & $152.4\pm 0.4$\\
    $L_*$ [\Lsun] & $1.7\pm 0.8$ \\
    $T_\text{eff}$ [K] & $4300\pm 100$\\
    $R_*$ [\rsun] & $2.3\pm 0.6$ \\
    $M_*$ [\msun] & $0.8\pm 0.1$ \\
    \prot\,[days] & $7.3\pm 0.2$ \\
    $R_\text{cor}$ [$R_*$] & $6.4\pm 1.7$ \\
    $B_\text{dip}$ [G] & $816$ \\
    $\Dot{M}_\text{acc}$ [\msunpyr] & $10^{-7}$ \\
    \hline\hline
    \end{tabular}
\end{table}

\section{Observations and data reduction}
\label{Sect:Observations}
\begin{table*}[t]
    \centering
    \caption{Log of the GRAVITY observations of S CrA.}
    \begin{tabular}{l l l l l l l l}
    \hline
    \hline
    HJD & Date & Time & Configuration & N & Seeing & $\tau_0$ & Calibrators\\
     &  & (UT) &  &  & ('') & (ms) & \\
    \hline
2459810.64606 & 2022-08-19 & 03:12 - 04:27 & UT1-UT2-UT3-UT4 & 9 & 0.6 - 0.8 & 8 - 11 & HD 186419\\
2459811.64606 & 2022-08-20 & 02:47 - 04:22 & UT1-UT2-UT3-UT4 & 12 & 0.8 - 1.1 & 6 - 10 & HD 176047, HD 186419\\
    \hline\hline
    \end{tabular}
    \label{tab:log}
\end{table*}

S CrA N was observed during two nights (August 19 and 20, 2022) as part of Guaranteed Time Observations with GRAVITY \citep{gravity_collaboration_first_2017} combining the four Unit Telescopes (UTs, 8.2~m in diameter each) of the ESO/VLTI. With interferometric baselines $B$ ranging between 40 and 130~m, we reached an angular resolution of $\lambda$/2$B_{\rm max}$~=~1.7~mas at a wavelength of $\lambda$~=~2.2~$\upmu$m, corresponding to 0.26 au at 152 pc. The instrument was used in dual field mode \citep[see][]{gravity_collaboration_first_2017} with S CrA S feeding the fringe tracker (FT) working at $\sim$ 1~kHz speed \citep{Lacour_FT} to freeze the atmospheric effects and lock the interferometric signals for the science channel (SC). Once the fringes are locked, we recorded $N$ 6-minute long sequences (12 exposures of 30~s) every $\sim$~10 minutes on S CrA N with the SC at high spectral resolution (R$\sim$4,000; $\Delta v = 74$ km/s). Nine files were recorded on August 19, and twelve on August 20. To calibrate the instrumental transfer function, we observed different calibrators sandwiching the exposures on the scientific target. They were chosen to be single stars, close in angular distance and magnitude to the target, and with very small and known diameters thanks to the SearchCal software \citep{SearchCal} of the JMMC\footnote{Available at www.jmmc.fr/searchcal}. For both nights, the atmospheric conditions were good with a seeing ranging from 0.6" to 1.1", and a coherence time $\tau_0$ longer that 6~ms. 
We give the detailed log of the observations in Table ~\ref{tab:log}.

We processed all the data with the standard GRAVITY pipeline \citep{gravity_DRS}. Each file contains the complex visibilities and closure phases of S CrA S in the FT, and of S CrA N in the SC. For S CrA N, the differential visibilities and phases as a function of the spectral channels, notably across the \Brgam~line, can be retrieved. When comparing the observations of the calibrator HD~186419 that is common to the two nights, we detected a chromatic effect in the SC data of the second night, as displayed in Fig. A.1. We thus calibrated all the SC data of the second night with the first calibrator, HD 176047, which does not show this effect. When looking at the S CrA N data, these chromatic effects are observed for most of the observations of the first night, and for a few files of the second night. Because such an effect might affect the absolute calibration of the visibility, we only used the data from the second night to constrain the dusty inner ring. This effect does not impact the differential observables across the \Brgam~line.

\section{Results}
\label{Sec:Results}
The data was analysed in two steps. First, a fitting of the whole K-band continuum emission was performed to constrain the dust distribution. Then, the Hydrogen emitting region was studied, 
as differential visibilities and differential phases were observed across the \Brgam~line.

\subsection{The K-band continuum emitting region}
\begin{figure*}[t]
    \centering
    \includegraphics[trim=0 0 0 0, clip, width=\linewidth]{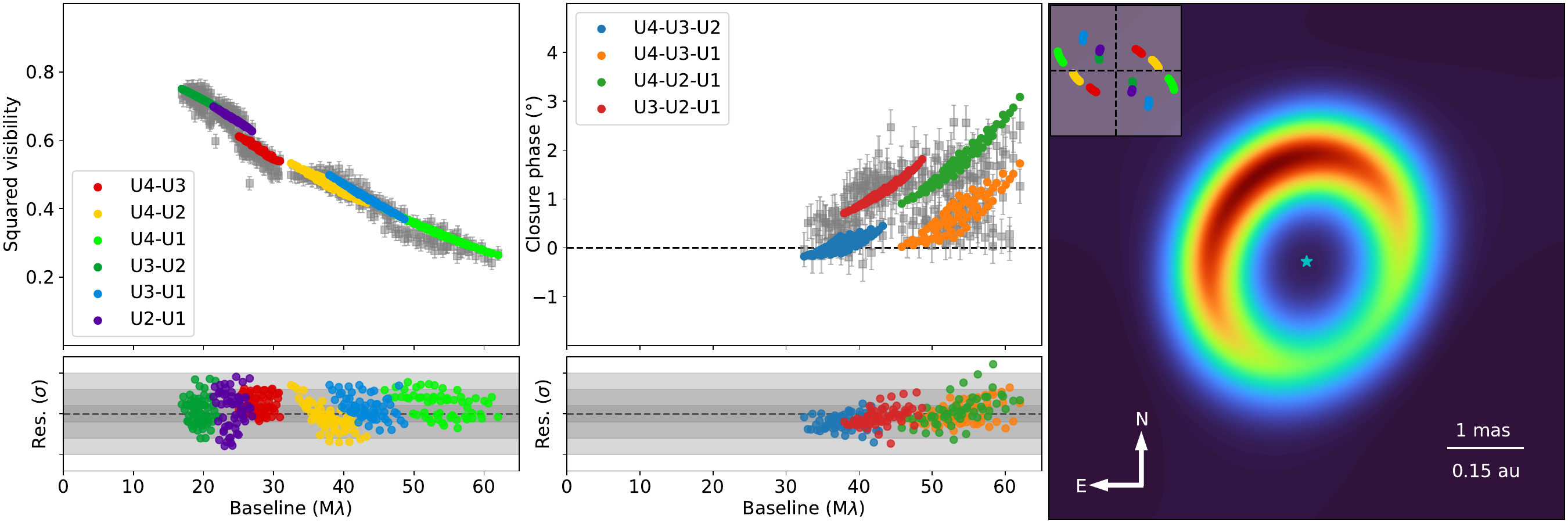}
    \caption{GRAVITY observations of S~CrA~N on 2022 August, 20: K-band continuum visibilities squared (left) and closure phases (center) as a function of baseline (grey symbols) superimposed with the best-fit ring model (right). The modeled quantities are depicted in colors matching the (u-v) plane coverage shown in the right-side inset. The lower panels show the residuals of the model, in units of $\sigma$. Shaded bands highlight 1-, 3-, and 5-$\sigma$ levels. }
    \label{fig:Continuum}
\end{figure*}

We used the visibilities squared and the closure phases over the K-band of the second night to constrain the geometry of the corresponding emitting region. We binned the SC data to obtain 5 spectral channels over the whole band.
The emitting region appears to be well resolved with visibilities squared as low as 0.25 at the longest baselines, and to exhibit a slight departure from centro-symmetry with closure phases of 1-2\ddeg~(Fig.~\ref{fig:Continuum}). We followed the same approach as \cite{lazareff_structure_2017} and \cite{gravity_2019,gravity_2021} to fit the visibility curve as a function of the baseline and the closure phase signals with a geometrical model. To consider conservative error bars, we fixed floor values on the error estimates provided by the data reduction pipeline, amounting to 2\% on the visibilities squared and to 0.35\ddeg~on the closure phases as estimated by the root mean square (rms) variations at the center of the K-band. We fitted our data with a composite model including a point-like source for the star, a halo contribution from the extended environment, and a first order azimuthally-modulated Gaussian ring component accounting for the emission of the inner dusty disk. At the spatial frequencies (u, v), the complex visibility is thus described by the linear combination:
\begin{equation}
    V_\text{Cont} (u, v) = \frac{F_* V_* (u, v) + F_\text{halo} V_\text{halo} (u, v) + F_\text{ring} V_\text{ring} (u, v)}{F_*+F_\text{halo}+F_\text{ring}},
\end{equation}
where $V_*$~=~1 is the visibility of the star that is unresolved at our angular resolution; $V_\text{halo}$~=~0 is the visibility of the component that is supposed to be much more extended than the GRAVITY field of view and fully resolved, even by the shortest baselines; $V_\text{ring}$ is the visibility of the ring model; $F_*$, $F_\text{halo}$, and $F_\text{ring}$ are the flux contributions of the star, the halo, and the ring, respectively, with $F_* + F_\text{halo} + F_\text{ring} = 1$. 

We refer to \citet{lazareff_structure_2017} for a detailed description of the modeling approach. More specifically, see their Table~5 for the analytical expression of $V_\text{ring} (u, v)$, their Table~7 for the parametrizations of the ring model, their equations (4) and (8) for the spectral dependence and the parametrization of the azimuthally-modulated ring ($c_1$ and $s_1$ coefficients), and their Sect. 3.3 for the fitting procedure. The spectral index of the star being fixed from its effective temperature, the nine free parameters of our geometrical ring model are the flux contributions of the halo and of the ring, the half-light radius of the Gaussian ring, its inclination and position angle, its width relative to its half-light radius, its spectral index, and the two coefficients $c_1$ and $s_1$ of the azimuthal modulation. Since the dusty environment is not fully resolved, its flux contribution and its size are partially degenerated as emphasized in \citet{lazareff_structure_2017}. We used a starting value of $F_\text{ring}$ between 0.35 and 0.45 estimated from the spectral energy distribution of \citet{Varga_2018} but kept this parameter free in the whole fitting process. \\
The parameters of the best-fit ring model are given in Table 3 and shown in Fig.~\ref{fig:Continuum}, along with their corresponding residuals, when comparing to the SC data. The best model corresponds to a Gaussian ring contributing to about 40\% of the total flux, with a half-flux radius of 1.55~mas (i.e., 0.24~au at 152.4~pc), an inclination of 30\ddeg, a position angle (PA) of about 150\ddeg, a spectral index range corresponding to a black-body emission at about 1800-2100~K, and an intensity enhancement located in the North-East side (see Fig.~\ref{fig:Continuum}-right).

\begin{table}[h]
\label{tab:ring}
        \caption{Best-fit parameters for the azimuthally modulated ring model. The Position Angle is from North to East. Inclination of 0\ddeg~corresponds to a face-on disk. The reported error bars are deduced from the uncertainty on the ring flux contribution.}
    \centering
    \begin{tabular}{l l l}
    \hline
    \hline
    Parameters [unit] & Best ring model & Description \\
    \hline
    $f_h$ [-] & $ 0.09 \pm 0.01$ & Fraction of halo flux \\
    $f_c$ [-] & $ 0.40 \pm 0.05$ & Fraction of ring flux   \\
    $i$ [\ddeg] & $ 30 \pm 5$& Inclination  \\
    $PA$ [\ddeg] & $ 151 \pm 5$ & Position Angle \\
    $R_K$ [mas] &  $ 1.55 \pm 0.15$& Half-light radius  \\
    $R_K$ [au] & $ 0.24 \pm 0.02$& Half-light radius \\
    $w$ [\%] & $ 40^{+17}_{-13}$& Relative width \\
    $k_d$ [-] & $-0.5 \pm 0.2$& Spectral index \\
    $c_1$ [-] & $-0.23 \pm 0.07$& Cosine modulation  \\
    $s_1$ [-] & $0.33 \pm 0.06$& Sine modulation \\
    \hline
    $\chi^2_r$ & 0.97&  \\
    \hline\hline
\end{tabular}
\end{table}

\subsection{The \Brgam~line spectrum}
\begin{figure*}
    \centering
    \includegraphics[trim=105 30 70 38,clip,width=0.95\linewidth]{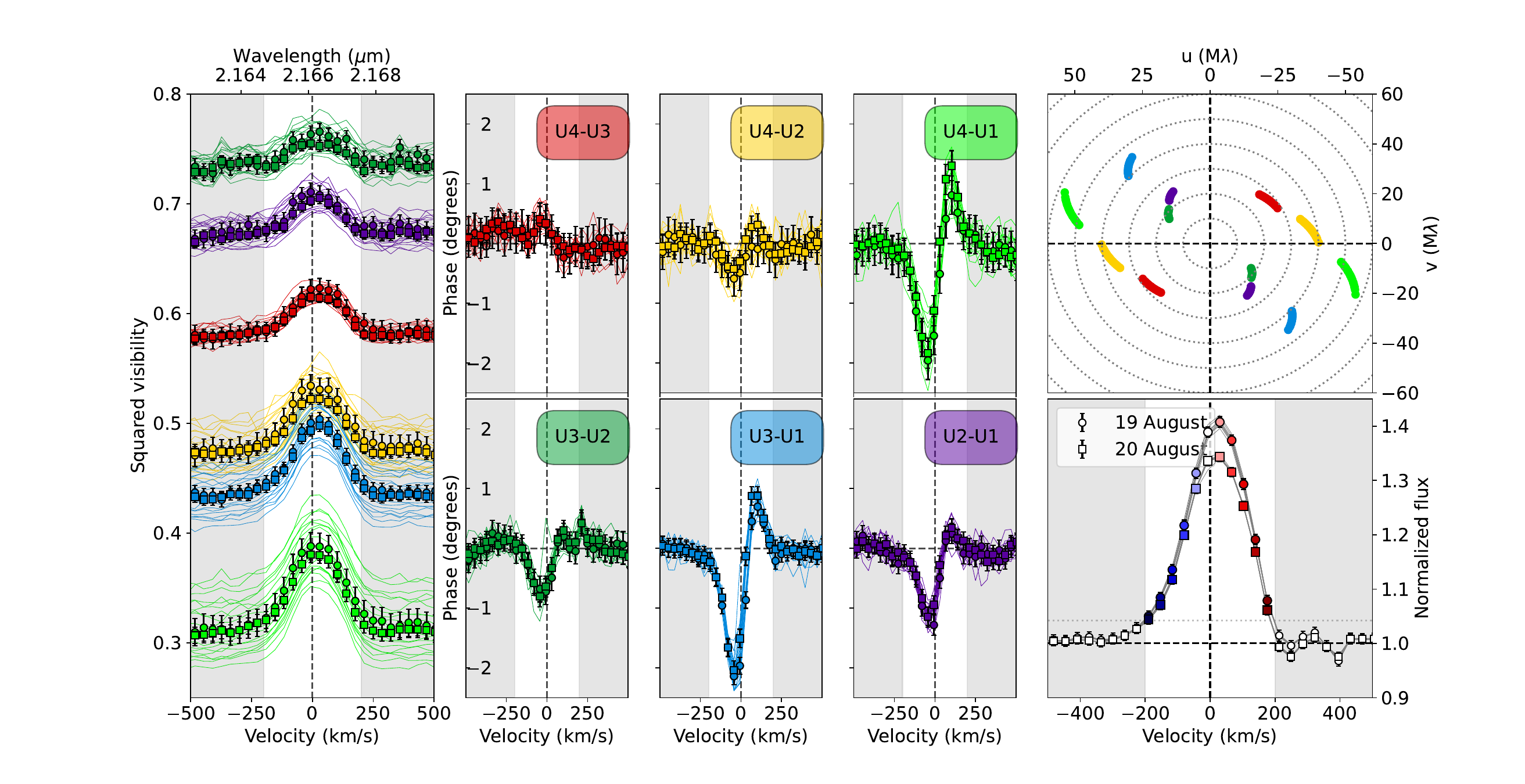}
    \caption{\Brgam~differential interferometric observables of S~CrA~N: Squared differential visibilities (left); Differential phases (middle); u-v plane coverage (upper-right); Line-to-continuum flux ratio in the reference frame of the star, LSR- tellurics-corrected (lower-right). Each color used in the u-v coverage plot corresponds to an interferometric baseline and is consistent from one panel to the other. For the interferometric observables, the plots include the quantities for each frame (solid lines), the median for August 19 (circles), and for August 20 (squares). Uncertainties are given in all plots (vertical bars), and are smaller than the symbol if not visible. Grey shaded regions correspond to the velocities that are not considered for our \Brgam~line analysis (see text for details).}
    \label{fig:BrGam_Data}
\end{figure*}

As GRAVITY is equipped with a spectrograph, we could retrieve the \Brgam~line spectrum for each observing date.
To this end, we first performed Local Standard of Rest (LSR) correction for all the observations, so the kinematics of the line are expressed in the frame of the star. Then, we used the PMOIRED python library \citep{merand_2022} to correct the GRAVITY wavelength calibration, based on a telluric model, and obtain a normalized spectrum corrected from this telluric spectrum (see Appendix~\ref{App:PMOIRED} for the detailed process).
No additional contribution of the photosphere to the \Brgam~line was considered, because the veiling of S CrA N is strong enough in the near-infrared \citep[$\sim$~4 in H-band;][]{Sullivan_2019} to consider the photospheric component negligible with respect to the line emission.
We are left with the \Brgam~line-to-continuum flux ratios $F_\text{L/C}$ for both dates, which are shown in the bottom-right panel of Fig.~\ref{fig:BrGam_Data}. This quantity shows very little dispersion during a single night over the whole line ($\leq$ 7\%). It also shows a 13\% variability between the two epochs, indicating a definite clear variation in the accretion-ejection processes on a daily timescale, which is expected for T Tauri stars \citep[e.g.][]{Hartmann_2016}. The line also shows a remarkable asymmetry in the profile, with a milder blue wing, and a steeper red wing with a peak slightly centered in the red (+30 km/s). 
We used $F_\text{L/C}$ to define the velocity domain where we consider the interferometric differential quantities across the \Brgam~line as relevant: we took all the spectral channels for which $F_\text{L/C}$ exceeds 10$\%$ of the peak value observed, 
which corresponds to a threshold of $F_\text{L/C}>1.04$ (or $\sim$ 40 for the signal-to-noise ratio (S/N)). The relevant spectral channels are illustrated with a blue-to-red color, and the threshold limit is delimited by a dotted horizontal grey line in Fig.~\ref{fig:BrGam_Data}.

\subsection{The morphology of the \Brgam~line emitting region}
We analyzed the differential interferometric observables to constrain the geometry of the regions responsible for the emission of the \Brgam~line in S CrA N. The remarkable S/N observed both in differential visibility (up to 40) and phase (up to 10) allow relative sizes and astrometric displacements to be measured with an accuracy of 20~\muas{} and 5~\muas, respectively.

\subsubsection{The differential interferometric data}
We 
forced the visibilities squared in the continuum adjacent to the Br$\gamma$ line to match the absolute visibility values predicted by our K-band continuum best-fit model. As depicted in Fig.~\ref{fig:BrGam_Data}-left, the differential visibilities are clearly higher across the emission line than in the continuum, indicating a more compact line emitting region than the $\sim 0.24$~au inner dusty disk ring model presented earlier. Over both nights, the scatter of the differential visibilities for the different frames is noticeable, notably for the longest baselines (in blue and green in Fig.~\ref{fig:BrGam_Data}-left) due to the change of the projected baselines during the night. 
Regarding the differential phases, a clear signal is detected for all baselines in the \Brgam~line individual frames, revealing an asymmetric emitting region. The higher amplitudes of a few degrees peak-to-peak are detected for the longest baselines, when the region is more resolved. The S-shape of the differential phases appears asymmetric for all the baselines, with a stronger amplitude in the blue part of the line. 

\subsubsection{The pure-line quantities}
To constrain the morphology of the \Brgam~line emitting region, we disentangled the contributions from the continuum and the line, using the line-to-continuum flux ratio $F_\text{L/C}$. 
In the regime of marginally resolved sources, we extracted the so-called pure-line visibility $V_\text{line}$ \citep[see Appendix B of][for the developments leading to this formula]{gravity_collaboration_gravity_2023}:
\begin{equation}
    V_\text{line}=\frac{F_\text{L/C} V_\text{tot} - V_\text{cont}}{F_\text{L/C} - 1},
    \label{eq:pureline_vis}
\end{equation}

The pure line visibility was used to derive the pure-line differential phases across the \Brgam~line $\phi_\text{line}$, which was obtained from the total phase $\phi_\text{tot}$ thanks to the total visibility $V_\text{tot}$, and the line-to-continuum flux ratio $F_\text{L/C}$ as follows \citep[see][for the details]{gravity_collaboration_gravity_2023}:
\begin{equation}
    \phi_\text{line} = \text{arcsin}\left( \frac{F_\text{L/C}}{F_\text{L/C}-1} \frac{V_\text{tot}}{V_\text{line}}\sin{\phi_\text{tot}}\right),
    \label{eq:pureline_phi}
\end{equation}
where $\phi_\text{tot}$ is the total differential phase (which is shown for each baseline in Fig~\ref{fig:BrGam_Data}). This quantity was computed for each observation and at each baseline, before being combined in order to get a single pure-line differential phase per night and per channel. Each value was taken as the median over all observations, and the associated uncertainties were computed following Student's t-distribution. The benefit of such a merge is a significant decrease in the uncertainties, resulting in a much more precise estimate of the line's photocenter displacements. The resulting pure-line phases are shown for each baseline in Fig.~\ref{fig:shifts}-left, superimposed to the individual frames' total phases.

\subsubsection{Asymmetry of the \Brgam~emitting region}

\begin{figure*}
    \centering
    \includegraphics[width=0.95\linewidth, trim =0 0 0 0, clip]{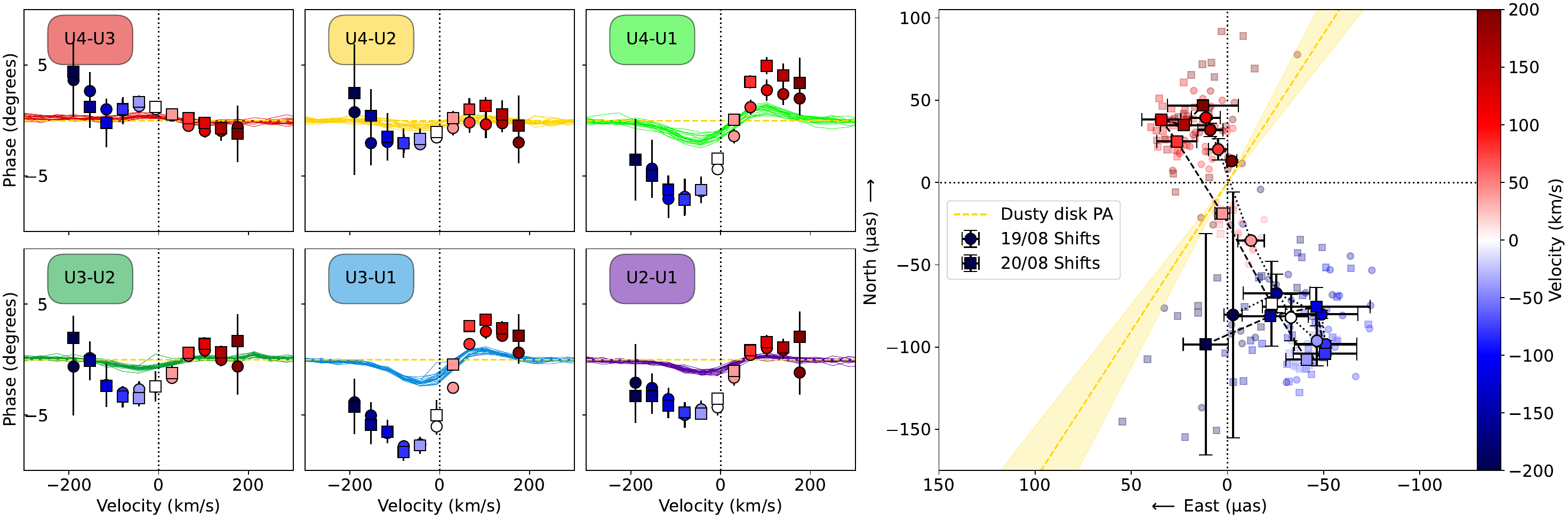}
    \caption{Asymmetry of the \Brgam~emitting region of S~CrA~N. Left : pure-line differential phases. The pure line differential phases are depicted by blue-to-red color coded symbols for August 19 (circles), and for August 20 (squares), while the observed phases for each individual observation (similar to Fig.~\ref{fig:BrGam_Data}) are depicted in solid lines, with their color corresponding to their baseline. Right : on-sky photocenter shifts with respect to the continuum photocenter (in (0, 0) as a function of velocity (coded from blue to red colors). 
    The light yellow cones correspond to the uncertainties on the major axis orientation of the inner dusty disk. The background symbols are the displacements obtained from individual files. }
    \label{fig:shifts}
\end{figure*}

To measure the on-sky photocenter shift across \Brgam, we used the linear relationship of marginally resolved sources between the photometric displacement's coordinate vector $\Vec{p}$ along a given projected baseline matrix $\Vec{B}$ and the associated pure-line differential phase matrix $\Vec{\phi_\textbf{line}}$ at a given wavelength $\lambda$ \citep{Lachaume_2003}: 
\begin{equation}
    \Vec{p}.\Vec{B} = - \lambda\;\frac{\Vec{\phi_\textbf{line}}}{2\pi}
    \label{eq:photocentershift}
\end{equation}
In the present case, $\Vec{B}$ and $\Vec{\phi_\textbf{line}}$ are a ($2\times 6$) and a ($1\times 6$) matrix, respectively. By inverting $\Vec{B}$, one recovers the coordinates of the photocenter shift $\Vec{p}$ relative to the photocenter of the continuum. We performed this inversion for all velocity channels of the \Brgam~line. This way, we obtained the astrometric displacement of \Brgam~for each spectral channel (Fig.~\ref{fig:shifts}-right), which is totally independent of a model. These photocenter shifts form a rather simple pattern, with a preferential orientation
misaligned by $\sim 45$\ddeg with respect to the {major axis PA} of the inner dusty disk. They lie in a very compact configuration (within $\sim$~150~\muas, i.e., $\sim$~0.023~au, which corresponds to a tenth of the half-flux radius of the K-band continuum emission.
The blue velocity channels (material moving toward the observer) centered around the null velocity (white circles/squares), and the red velocity channels (material moving away from the observer) clearly depart from the null velocity in the North-East direction for $\sim 150$~\muas~until $v\sim100$ km/s. Beyond this velocity, the shifts remain centered around the same position. The position of the star relative to the continuum's photocenter is consistent with the null velocity shifts, when propagating the uncertainties of the continuum model : its coordinates in the reference frame attached to the photocenter of the continuum are $-102\pm95$ \muas{} towards East and $-164\pm95$ \muas{} towards North. It is worth noticing that despite they are affected by larger error bars, the individual files' displacements (visible as light colored background symbols in Fig.~\ref{fig:shifts}) seem to form a more complex pattern with departures of the higher velocities from the straight line mentioned here. 

\subsubsection{Size of the \Brgam~emitting region}

\begin{figure}[h]
    \centering
    \includegraphics[trim= 5 60 15 80, clip, width=0.8\linewidth]{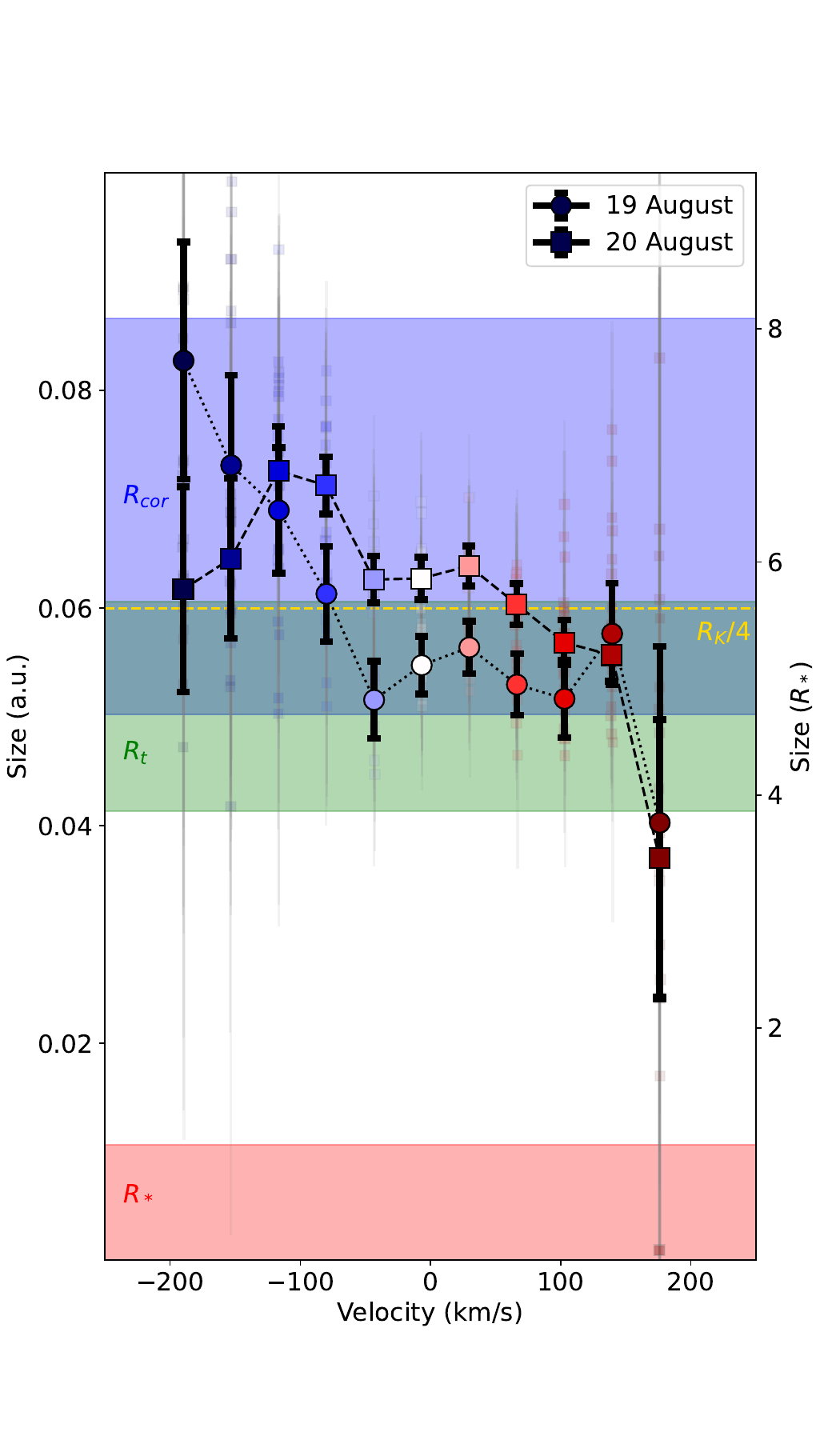}
    \caption{Characteristic sizes of the \Brgam~emitting region of S~CrA~N as a function of velocity (coded from blue to red colors) for August 19 (circles), and for August 20 (squares). The red, green and blue areas correspond to the stellar, the truncation, and the corotation radii, respectively. The dashed gold line marks the fourth of the half-light radius derived for the inner dusty ring model, for the sake of comparison. Colored symbols with grey error bars in the background correspond to the sizes derived from individual files. The distance used is 154.2 pc.}
    \label{fig:sizes}
\end{figure}

\begin{figure}[t]
    \centering
    \includegraphics[trim=380 0 0 0,clip,width=\linewidth]{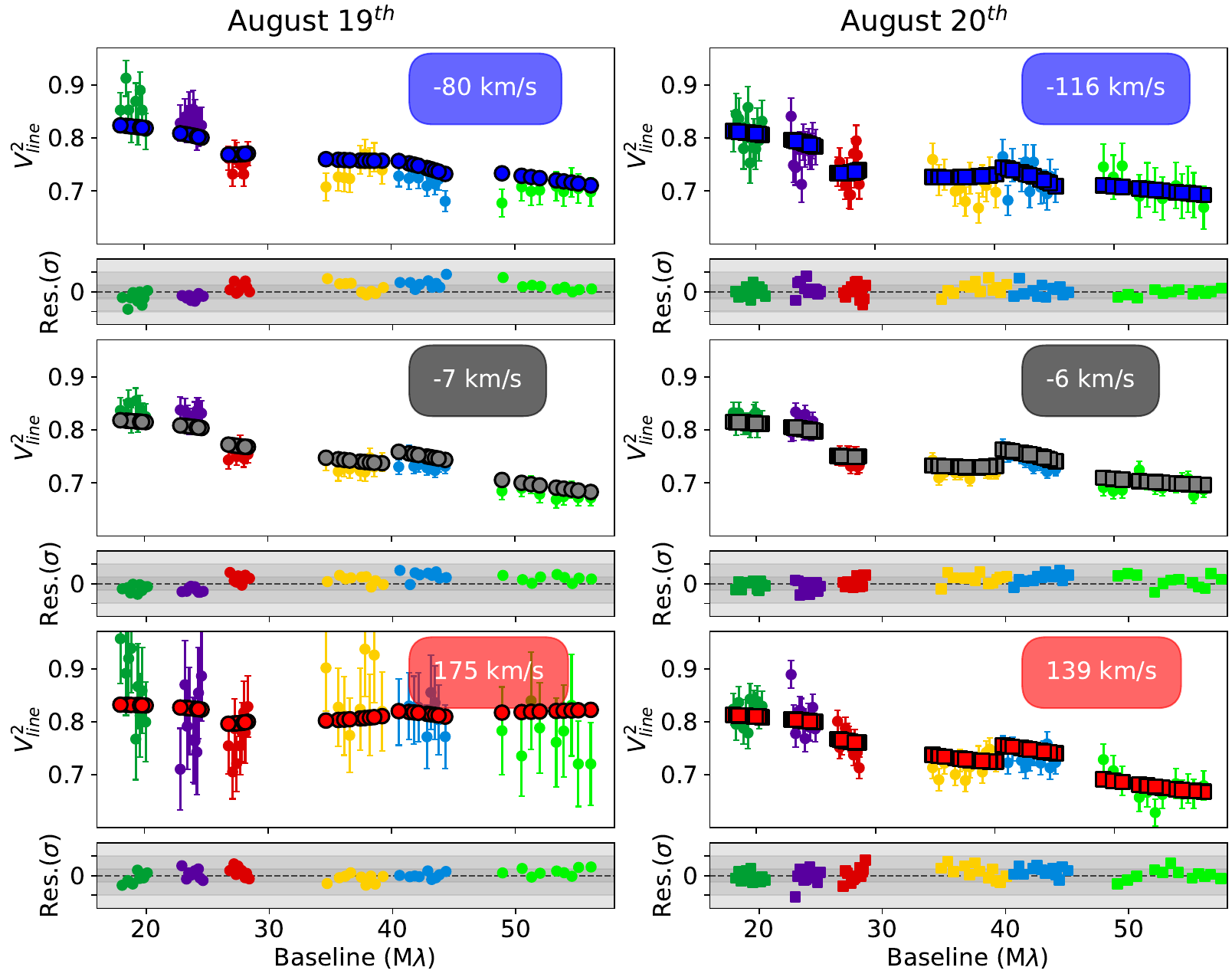}
    \caption{Pure line visibility fits for August 19\Th (left column) and 20\Th (right column). Each row consists of two subplots : The top one includes the pure line visibilities squared and their 1-$\sigma$ uncertainty observed at a given velocity (mentioned in the panel), superimposed with the corresponding model including a Gaussian disk and a halo. The bottom one illustrates the residuals of the model, in units of $\sigma$. Three shades of grey mark the 1-,3-, and 5-$\sigma$ levels. The color of the data points codes for the baseline.}
    \label{fig:sizes-fit}
\end{figure}

A way to disentangle between the different processes at play, and go deeper in the understanding of the morphology of the emitting region is to use a geometrical model to reproduce the spatial distribution of the emitting region at each velocity channel of the \Brgam~line. In the case of S~CrA~N, the pure-line visibilities (especially at low frequencies) are not satisfyingly reproduced by a Gaussian disk alone \citep[as it is usually done for CTTS, see e.g.][]{gravity_2023_CITau, gravity_collaboration_gravity_2023}, so a composite model combining a Gaussian disk model and a fully resolved halo was adjusted to the data. The expression of the pure-line visibility is then (see Appendix~\ref{App:halo} for the justifications and developments leading to the formula):
\begin{equation}
    V_\text{line} = (1 - C_\text{H})\,\exp{\left(-\frac{\pi^2B^2\,\Theta^2}{\lambda^2\ln 2}\right)}.
\end{equation}
$C_\text{H}$ is the relative contribution of the halo to the total flux, $\Theta$ is the half-width at half-maximum (HWHM) of the Gaussian disk, $B$ and $\lambda$ are a given projected baseline and wavelength of observation. This projected baseline takes into account an inclination of the disk ($\alpha$) along a position angle ($PA$) measured from North to East. This model is adjusted independently to the nine observations of August 19, and to the 12 observations of August 20, taking full advantage of the high quality of the single frames data to provide strong constraints to the $\chi^2$ minimization procedure. Because a degeneracy is expected between the halo and the size (in a marginally resolved regime), the data were fitted first with a contribution of the halo fixed to the value of the continuum (10\%). Then, the contribution of the halo was adjusted, while the other parameters were set to their values minimizing $\chi^2$ in the previous procedure. The best model corresponds to a disk whose HWHM equals $0.43\pm 0.06$ mas ($0.06\pm 0.01$ au), inclined by 51$^\circ \pm 13$\ddeg~around a position angle of $141^\circ \pm 10^\circ$. The contribution of the halo is within $8.9~\pm~0.3\%$. These models yield $\chi^2_r$ of 0.83 on average, with residuals contained within 1-$\sigma$ for the bulk of points.
The best-fit models for all velocity channels are illustrated in Fig.~\ref{fig:sizes}, they are detailed in Table~\ref{tab:disk}, and six of them are compared against the measured pure line visibilities in Fig.~\ref{fig:sizes-fit}. 
\begin{table*}[t]
        \caption{Best-fit parameters with 1$\sigma$ error bars for the Halo + Gaussian disk model of the \Brgam~pure-line visibilities. The position angles are from North to East.}
    \centering
    \begin{tabular}{c | c c c c c | c c c c c}
    \hline
    \hline
    $ $ & \multicolumn{5}{c|}{19 August} & \multicolumn{5}{c}{20 August} \\
    \hline
    Velocity & HWHM & $\alpha$ & $PA$ & $C_\text{H}$ & $\chi_r^2$ & HWHM & $\alpha$ & $PA$ & $C_\text{H}$ & $\chi_r^2$ \\
    $\left[\right.$km/s] & [mas] & [\ddeg] & [\ddeg] & [$\%$] & [-] & [mas] & [\ddeg] & [\ddeg] & [$\%$] & [-] \\
    \hline
    -190 & $0.54\pm0.07$ & $54\pm8$ & $142\pm14$ & $9.6\pm2.7$ & 0.44 & $0.40\pm 0.06$ & $44\pm 12$ & $131\pm 28$ & $9.6\pm 2.1$ & 0.56 \\
    -153 & $0.48\pm0.05$ & $54\pm7$ & $153\pm8$ & $8.7\pm1.5$ & 0.47 & $0.42\pm 0.05$ & $48\pm 8$ & $150\pm 10$ & $9.3\pm 1.4$ & 0.69 \\
    -117 & $0.45\pm0.04$ & $52\pm5$ & $157\pm5$ & $8.7\pm0.9$ & 0.67 & $0.48\pm0.03$ & $56\pm 4$ & $153\pm 4$ & $9.0\pm 0.8$ & 0.79 \\
    -80 & $0.40\pm0.03$ & $48\pm5$ & $159\pm5$ & $8.7\pm0.5$ & 1.15 & $0.47\pm0.02$ & $55\pm 2$ & $151\pm 2$ & $8.7\pm 0.5$ & 1.30 \\
    -44 & $0.34\pm0.02$ & $37\pm6$ & $149\pm11$ & $8.7\pm0.3$ & 0.97 & $0.41\pm0.01$ & $52\pm 2$ & $143\pm 3$ & $9.0\pm 0.3$ & 0.97 \\
    -7 & $0.36\pm0.02$ & $41\pm4$ & $141\pm9$ & $8.7\pm0.3$ & 0.88 & $0.41\pm0.01$ & $52\pm 2$ & $142\pm 3$ & $9.0\pm 0.2$ & 1.26 \\
    +30 & $0.37\pm0.02$ & $45\pm3$ & $137\pm7$ & $8.7\pm0.5$ & 0.64 & $0.42\pm0.01$ & $53\pm 2$ & $139\pm 3$ & $9.0\pm 0.3$ & 1.04 \\
    +66 & $0.35\pm0.02$ & $37\pm4$ & $144\pm11$ & $9.0\pm0.5$ & 0.56 & $0.40\pm0.01$ & $52\pm 2$ & $135\pm 4$ & $8.7\pm 0.3$ & 1.03 \\
    +103 & $0.34\pm0.02$ & $31\pm7$ & $148\pm17$ & $9.0\pm0.5$ & 0.56 & $0.37\pm0.01$ & $48\pm 2$ & $130\pm 6$ & $9.0\pm 0.3$ & 0.76 \\
    +139 & $0.38\pm 0.03$ & $39\pm7$ & $149\pm 12$ & $8.7\pm 0.6$ & 0.96 & $0.37\pm 0.02$ & $45\pm4$ & $122\pm 10$ & $9.0\pm0.5$ & 1.22 \\
    +176 & $0.26\pm0.11$ & $90\pm90$ & $156\pm19$ & $8.7\pm1.7$ & 0.62 & $0.24\pm 0.08$ & $90\pm 90$ & $137\pm 26$ & $9.0\pm 1.7$ & 0.78 \\
    \hline\hline
\end{tabular}
\label{tab:disk}
\end{table*}

\section{Discussion}
\label{Sec:Discussion}

From our VLTI/GRAVITY observations obtained on S~CrA~N on two successive nights of August 2022, we derived the characteristic sizes of the near-infrared emitting regions and traced their on-sky displacements. The outstanding quality of the data sets provides new insight into the innermost regions of this strong accretor. In this section, we discuss how these results constrain the accretion-ejection processes in a global view of this young system.

\subsection{The star-disk interaction region}

It is generally accepted that the star-disk interaction for T Tauri stars is dominated by magnetospheric accretion: the $\sim$1 kG stellar magnetic field truncates the inner gaseous disk at the truncation radius $R_t$; the matter is accreted onto the central star along the magnetic funnels and falls on a accretion spot. Combined with a heated inner disk in Keplerian rotation, this scenario offers several origins for the \Brgam~emission line. Our clear astrometric offset of the \Brgam~emitting region of about $\sim$~150~\muas~(i.e., 0.023 au) that is misaligned by $\sim 45$\ddeg~with respect to the major axis of the inner dusty disk (Fig.~\ref{fig:shifts}) does not favor a pure disk in Keplerian rotation, for which the photocenter shift would trace a line parallel to the disk's major axis, with the photocenters of the fastest velocities closer to the null-velocity than the photocenters of the slowest velocities \citep[see][for a typical example]{Mendigutia}. Instead, our astrometric displacements are more in agreement with polar phenomena, e.g. accretion flows and/or outflows. The visibilities recorded with GRAVITY in the \Brgam~line correspond to an emitting region of $\sim$0.05-0.06~au at the zero-velocity channel, which is located well within the inner dusty disk for which we derived a half-flux radius of 0.24~au. This characteristic size of the \Brgam~line region corresponds to $\sim$5~$R_*$, which is close to the truncation radius $R_t=4.4\pm 0.9\,R_*$ ($\sim$~0.05~au) we obtained when applying the scaling of Pantolmos et al. (in prep) with the S~CrA~N parameters given in Table~\ref{tab:param}. As depicted in Fig.~\ref{fig:sizes}, some blueshifted channels do extend further than the truncation radius, but all are included within the corotation radius $R_{\rm cor}=6.4\pm1.7~R_*$ ($\sim$~0.07~au). 
Noticing that the Gaussian model we used to model the \Brgam~line emitting region might underestimate the size of the magnetosphere by up to a factor two depending on the phase of observation \citep[see Fig. 6 of][]{tessore_2023}, the distribution of sizes we derived is definitely larger than $R_t$ in most of the spectral channels. This suggests that the \Brgam~is the result of a combination of magnetospheric accretion and outflow processes. Previous observations of blueshifted forbidden emission lines and P Cygni profiles in S~CrA~N support the presence of outflows in the vicinity of the star \citep[e.g.][N23]{carmona_2007, gahm_2018}. This is also supported by the comparison of our findings with Magneto-Hydro-Dynamic (MHD) simulations. Wojtczak et al. (in revision) computed synthetic interferometric observables for the strongly accreting T Tauri star RU~Lup, with stellar parameters and inclination similar to those of S~CrA~N ($M_*$ = 0.8~\msun; $R_*$ = 2.5~\rsun; $T_\text{eff}$ = 4300~K; \prot\ = 7 days; $\Dot{M}_\text{acc}$ = 2$\times 10^{-7}$\msunpyr; $d$ =  157.5~pc; $i\sim$~20$^\circ$, see Tables 1 and 3). Their hybrid model including an axisymmetric magnetosphere and a biconical magneto-centrifugal disk wind exhibits a constant size of about 0.06~au at the center of the \Brgam~line, a larger size in the blue wing (up to $\sim 0.1$~au), and astrometric displacements at the core of the line along a line of about 0.03~au misaligned with the inner disk major axis. All these characteristics are remarkably similar to our observations (see their Fig. 7-left). When comparing to the other targets of the GRAVITY sample \citep{gravity_2021, gravity_collaboration_gravity_2023}, S~CrA~N appears to be similar with the strongly accreting ones (e.g. RU Lup, AS 205 N), with a sublimation radius located at 0.1-0.2~au, and a \Brgam~emission more extended than the corotation radius, suggesting an origin from a combination of magnetospheric accretion and disk winds.

\subsection{Evidence for sustained disk winds}

Beyond the truncation radius, the open magnetic field lines of the star are expected to produce transient magnetospheric ejections as well as disk winds \citep[see, e.g.][]{romanova_2009,zanni_2013,pantolmos_2020}. In their spectroscopic study of S~CrA~N, N23 detected outflows' signatures (namely Balmer lines blueshifted absorptions, [OI] and [SII] forbidden emissions) that might be reproduced by a hot ($\sim 9000$~K), dense ($\sim 5\times 10^{-11}$ g.cm$^{-3}$) wind launched from a region extending up to $\sim 20$ $R_*$ (i.e., 0.24~au), with a mass loss rate of $\Dot{M}_{\rm wind}\sim 10^{-8}$\msunpyr. This size is comparable to the range of inner rim's location (0.13-0.24~au) determined for the dusty disk in the K-band \citep[see][and this work]{vural_2012,gravity_collaboration_wind_2017}. A good tracer of the kinematics and structure of the inner gaseous disk is the $^{12}$CO gas emission \citep{pontoppidan_2011}. Using CRIRES(+) observations of S~CrA~N, \citet{banzatti_2022} and \citet{Grant_2024} detected triangular CO lines in emission and two distinct velocity components of the CO emission: the broad (BC) and the narrow (NC) components. When scaling their values to our stellar parameters (Table~\ref{tab:param}), we derive a half-width at half maximum for BC, $R_{\rm CO}^{\rm BC}$, of $0.4\pm0.1$~au for \citet{banzatti_2022} and of $0.14\pm0.05$ au for \cite{Grant_2024}, respectively; for NC, $R_{\rm CO}^{\rm NC}$ of $8\pm3$ au and $4\pm1$, respectively. When comparing with the half-flux radius of the inner dusty disk (0.24~au), the CO broad component seems to originate from a region inside or close to the dust sublimation radius, while the NC component definitely arises from a region beyond the dust sublimation radius in the disk. To reproduce these CO features (shapes and radii of the emission), \citet{pontoppidan_2011} and \cite{banzatti_2022} have developed different models that support the presence of sustained disk winds with $\Dot{M}_{\rm wind}\sim 10^{-8}$\msunpyr, i.e., typically a tenth of $\Dot{M}_{\rm acc}$.

\subsection{Magnetospheric accretion as a dust-heating mechanism}

The near-infrared excess observed in the spectral energy distribution of CTTS is generally interpreted as the hot dust emission located at the inner edge of the disk where the dust grains sublimate. This sublimation radius (in au) can be computed by \citep{Monnier2002}:
\begin{equation}
    R_{\rm sub} = 1.1 \sqrt{Q_R} \sqrt{\frac{L_*}{1000 L_\odot}} \left( \frac{1500}{T_{\rm sub}} \right)^2,
    \label{eq:sub_rad}
\end{equation}
with $Q_R$ the absorption efficiencies of the dust, L$_*$ the luminosity of the star, and T$_{\rm sub}$ the temperature of dust sublimation.\\
Considering \te~=~4300~K for S~CrA~N and grain sizes ranging from 0.01~$\upmu$m to 1~$\upmu$m, $Q_R$ spreads between 1 and 4 (see their Fig.~2). With a typical sublimation temperature T$_{\rm sub}$ ranging between 1300~K and 1700~K and L$_*=1.7~L_\odot$, R$_{\rm sub}$ ranges between 0.04 and 0.15~au, which is twice less than the half-light radius of the dusty disk we derived (0.24~au). Our geometrical model is likely too simple to catch the complex radial and vertical structure of the inner rim but this might not explain the size difference. As \cite{Pinte_2008} pointed out that the stellar scattered light has a non-negligible contribution in the interferometric measurements for T Tauri stars (see their Fig. 3), we have added a halo in our model to account for this contribution, at least partly, and derived a halo contribution of $\sim$10\% in the K-band continuum and in the \Brgam~line. As this scattering is very unlikely to account for the large continuum K-band half-flux radius, additional dust heating sources and/or properties should be invoked. Indeed, the sublimation rim position strongly depends on the dust composition (pure silicate, here), cooling efficiency, and backwarming effects. More constraints are needed for a better description of the disk mineralogy (for instance in the mid-infrared with MATISSE).

As S~CrA~N is a strongly accreting T Tauri Star, we tested the heating by the stellar accretion spot(s). The spot(s) created by magnetospheric accretion at the surface of the star could be modeled as a $\sim$9000~K blackbody \citep[using Eq. 9 from][assuming a filling factor of 0.05 for the spot]{Hartmann_2016} which makes it a good candidate, especially since the previous spectropolarimetric study of N23 revealed the presence of such a feature. Even though the complete calculations of the intricate chemicophysical reactions of the dust in this context go beyond the scope of this paper, we can estimate the magnitude of change in $R_{\rm sub}$ provoked by the accretion shock heating. In Eq.~\ref{eq:sub_rad}, the stellar luminosity could be replaced by the sum of the stellar and accretion luminosities, the latter being computed assuming a complete conversion of kinetic energy into radiation, by :
\begin{equation}
    L_{\rm acc} = \,\frac{G\Dot{M}_{\rm acc}M_*}{R_*}\left(1-\frac{R_*}{R_t}\right)
    \label{eq:Lacc}
\end{equation}
Using the parameters from Table~\ref{tab:param} and a truncation radius $R_t=4.4~\pm$~0.9~R$_*$, we get $L_{\rm acc}$ = 0.9 $\pm$ 0.3~L$_\odot$. In this situation, an equivalent effective temperature can be obtained as $T_{\rm eq}^4=(1-f) T_{\rm eff}^4+f T_{\rm acc}^4$, where $f$ is the filling factor of the accretion spot. 
Keeping the assumption of $f=0.05$, $T_{\rm eq}\sim 5000$~K. Then, $Q_R$ ranges between 1 and 6 and consequently, $R_{\rm sub}$ increases up to 0.19~au, assuming the same kind of dust as previously.\\ 
As expected, the heating contribution from an accretion spot increases the sublimation radius. The remaining difference still leads to consider different contributions beyond this simple model. For example, an extended magnetosphere composed of one -or more- accretion column could contribute to the heating of the dust. A gaseous disk undergoing the strong accretion rates measured in S~CrA~N(10$^{-7}$\msunpyr), and therefore producing viscous heating can also be contemplated. If the former could be modeled \citep[as in, e.g.][]{Pittman_2022}, the latter is more difficult to constrain \textit{a priori}. Since no evidence of protoplanets have been found in the inner regions of this system so far, the presence of a cavity carved by a nascent planet appears more unlikely. 

\begin{figure*}[ht]
    \centering
    \includegraphics[trim=0 0 0 0, clip, width=\linewidth]{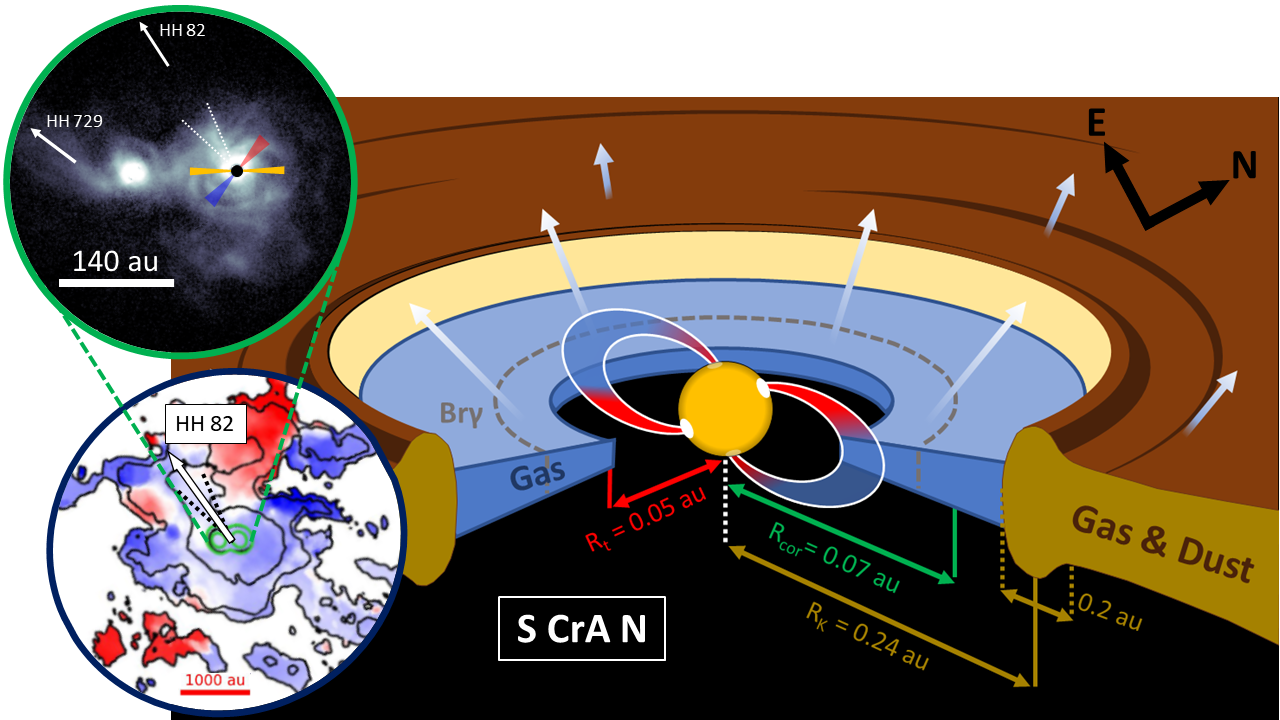}
    \caption{S~CrA~N at different scales. Right: Representation of the star-disk interaction region (not to scale), with the characteristic sizes derived in this work. {The obliquity of the magnetosphere is just illustrative as no consolidated value exists}. Top-left: SPHERE polarized intensity image from \citet{zhang_2023}, and the directions toward the HH objects associated to S~CrA. We also reported the PA of the dusty disk measured from the K-band continuum and the orientation of photocenter shifts as yellow, red and blue cones, respectively. Dashed cones highlight the suggested outflow emission (see text). Bottom-left: ALMA $^{12}$CO(2-1) first moment map from \citet{gupta_2023}. The cone is reported from the SPHERE observations, and the HH82 direction is indicated. Both images have been rotated to have the same orientation as the sketch, where the North and East directions are shown.}
    \label{fig:schema_scales}
\end{figure*}

\subsection{Hints of variable accretion and/or ejection ?}

Temporal variability in photometry and/or spectroscopy is frequently observed on different timescales for CTTS \citep{fischer_2023}. For S~CrA~N, previous CO observations \citep{banzatti_2022,Grant_2024} suggest a variable emission, and potentially a variable disk wind (see Sect. 4.2). Regarding the magnetospheric accretion process, the ratio of truncation to corotation radius $R_t/R_{\rm cor} = 0.7 \pm 0.2$ we obtained is close to the transition between stable and unstable accretion regime of 0.8 stated by Pantolmos et al. (in prep). This suggests a variable accretion process and supports the presence of two opposite accretion columns with different densities proposed by N23, as a consequence of the dipolar magnetic field and its non-null ($\geq$ 5$^\circ$) magnetic obliquity they derived. Thanks to the high quality of our GRAVITY data, we can investigate the variations across the \Brgam~line between the two nights of observation. Notably, we observed a flux decrease by 13\% between the two nights (Fig.\ref{fig:BrGam_Data}-bottom-right). Following the methodology of \citet{prato_2003} we measured the Equivalent Width (EW) of the \Brgam~line of our GRAVITY median spectra to probe the accretion rate between -500 and +500 km/s. We obtained -91$\pm$2~km/s and -82$\pm$1~km/s for August 19 and August 20, respectively, which corresponds to an average value of EW encountered in S~CrA~N for \Brgam{} \citep[between -30 km/s and -120 km/s, see][]{prato_2003, Sullivan_2019,gravity_collaboration_gravity_2023}. The EW determined from GRAVITY decreases between the two nights, suggesting that the star may exhibit variable accretion and/or variable ejection. Between the two nights, the interferometric signatures remain similar within our uncertainties (see Fig. 4 for the size of the \Brgam~line emitting region, and Fig. 3 for the on-sky displacements across the emission line). Based on the simulations of interferometric signatures for non-axisymmetric accreting magnetospheres as a function of rotation phase \citep[see Fig. 5 in][]{tessore_2023}, we would have expected to detect an orientation change of the on-sky displacements across the \Brgam~line in an interval of 1 day as the star (and the accretion funnel with it) has rotated by 2~$\pi$/7.3~rad. Since the photocenter shift's orientation remains clearly the same between the two nights, we thus favor a significant contribution from a more extended, axisymmetric component, likely a variable disk wind that could be the dominating origin of the \Brgam~line. Such a wind can be quite axisymmetric, preserving the photocenters and size of the region, and varies on the scale of a few days in terms of flux, if it is driven by gas dynamics at the magnetosphere-disk boundary where the Keplerian period is similar to the rotation period of the star \citep[see e.g. the models of][]{knigge_1995,lima_2010,zanni_2013}.


\subsection{The larger scale environment of S~CrA~N} 

Previous large-scale observations of the S~CrA system indicate the presence of structures at spatial scales ranging from 10 au to 10000 au that we can attempt to link to those detected in the innermost regions of S~CrA~N as depicted in Fig.~\ref{fig:schema_scales}. Regarding the dust distribution, SPHERE images in polarized light obtained by \citet{zhang_2023} reveal a complex environment around the S~CrA system with fragmented structures like spirals in the disk surrounding S~CrA~N at a few tens of au, and streamers around the secondary component (see Fig.~\ref{fig:schema_scales}-top). Due to the high complexity of the environment, the precise morphology of the disk around S~CrA~N is hard to constrain from these observations, and dust continuum emission is unresolved with ALMA \cite{cazzoletti_2019}, which would have been helpful to bridge the inner and the outer scales. While the polarized intensity images suggest a near side of the outer disk towards East \citep{zhang_2023}, our continuum best-fit model leads to a brighter and thus a far side in the East direction. Should the outer disk's orientation from SPHERE be confirmed, this would indicate a misalignment between the inner and the outer parts, which could lead to shadows in the disk that have not be detected so far. In the SPHERE images, a bright emission can be distinguished as a polar departure from S~CrA~N in the East direction (see dotted cone in Fig.~\ref{fig:schema_scales}-top). If real, this structure is remarkably well aligned toward HH~82, and could be interpreted as an outflow from the large-scale disk; similarly, the streamer from S~CrA~S in the South-East direction could be connected to HH~729, both Herbig-Haro objects being located at $\sim$20,000~au \citep{reipurth_1988,kumar_2011,peterson_2011}. Nevertheless, the position angles of these Herbig-Haro objects ($\sim$95\ddeg~and $\sim$115\ddeg) do not allow us to connect them to any of the phenomena discussed in our work so far. By the same way, regarding the gas distribution, ALMA observations exhibit a complex $^{12}$CO emission at a scale of a few thousands au with different kinematic features \citep[see Fig.~\ref{fig:schema_scales}-bottom;][]{gupta_2023}, and a blueshifted (by a few km/s) structure oriented toward HH~82 that do not seem to be connected with our observations. However the complex environment of the S~CrA system makes the existence of all these structures highly speculative, and new observations at higher resolution and/or intermediate scales (e.g. with VLTI/MATISSE) are required to better constrain the morphology of this complex system. 

\section{Conclusions}
\label{Sec:Conclusions}

In the present work, we used two successive VLTI/GRAVITY near-infrared interferometric observations with the UTs to constrain the innermost regions of the strong accretor S~CrA~N. For the first time, we managed to detect accurate interferometric differential signals in the \Brgam~emitting line with a cadence of about 10 minutes. We summarize our findings below:
\begin{itemize}
\item The K band continuum emission is reproduced with a first order azimuthally-modulated ring with a half-light radius of 0.24~au (20~$R_*$), an inclination of $\sim$30\ddeg, a position angle of $\sim$150\ddeg. The location of the brigthness' modulation suggests a near-side of the inner disk in the South-West direction. The half-flux radius appears to be variable over a few years, with previous determinations of 0.13-0.15~au in the K band. 
\item  We find that the sublimation radius should be at least twice closer than the half-light radius of the continuum we measure. This suggests that magnetospheric accretion might be an important mechanism responsible for dust sublimation on top of stellar luminosity. Beyond our simple approach of a single spot, the contribution of an extended magnetosphere and a hot inner viscous disk can also be envisioned, particularly for such a strong accretion rate ($\Dot{M}_{\rm acc} \sim 10^{-7}$ \msunpyr). 
\item A compact region is found to be responsible for the \Brgam{} emission ($\sim$0.06~au or 5-6~$R_*$). This emission presumably comes from both magnetospheric accretion and disk winds, based on the comparison of its size to the magnetic truncation radius. The on-sky displacements across the spectral channels of the \Brgam~line appear to be misaligned with respect to the dusty disk by $\sim$45\ddeg and extend up to 0.023~au. This misalignement suggests that these photocentershifts trace polar processes as magnetospheric accretion funnels and/or outflows. This is in line with the multiple evidence for disk winds in this source, notably spectroscopic signatures, and with the tilted dipolar magnetic field topology that has been reconstructed in a previous work.
\item We find this object to be in line with other strong accretors from the GRAVITY sample with an average sublimation radius (0.1-0.2~au), and a \Brgam~emitting region larger than the corotation radius, suggesting an origin for this emission from a combination of magnetospheric accretion and disk winds.
\item While the flux of the \Brgam~line moderately decreases from one night to the next by 13\%, the interferometric quantities remain almost unchanged, which favors a dominant contribution from an axisymmetric component like a disk wind.
\item We could not find any clear connection between the inner scales we probed with GRAVITY and the speculative complex dust and gas structures detected with SPHERE and ALMA.
\end{itemize}

These results demonstrate the power of near-infrared optical long-baseline interferometry in providing unique inputs for advanced modeling of acretion-ejection models and to probe the innermost regions of the young stellar systems by combining spatial, spectral, and temporal resolutions.

\begin{acknowledgements}
This work was supported by CNRS/INSU, by the ”Programme National de Physique Stellaire” (PNPS) of CNRS/INSU co-funded by CEA and CNES, and by Action Spécifique ASHRA of CNRS/INSU co-funded by CNES.\\
This work has been supported by the French National Research Agency (ANR) in the framework of the "Investissements d'Avenir" program (ANR-15-IDEX-02) and in the framework of the "ANR-23-EDIR-0001-01" project. \\
M.B. received funding from the European Research Council (ERC) under the European Union’s Horizon 2020 research and innovation programme (PROTOPLANETS, grant agreement No. 101002188). \\
This project has received funding from the European Research Council (ERC) under the European Union’s Horizon 2020 research and innovation programme (grant agree ment No742095;SPIDI: Star-Planets-Inner Disk-Interactions, http://www.spidi-eu.org). \\
A.C.G. acknowledges support from PRIN-MUR 2022 20228JPA3A “The path to star and planet formation in the JWST era (PATH)” funded by NextGeneration EU and by INAF-GoG 2022 “NIR-dark Accretion Outbursts in Massive Young stellar objects (NAOMY)” and Large Grant INAF 2022 “YSOs Outflows, Disks and Accretion: towards a global framework for the evolution of planet forming systems (YODA)”.\\
J.S.-B. acknowledges the support received from the UNAM PAPIIT project IA 105023\\
This research has made use of the Jean-Marie Mariotti Center \texttt{Aspro}\footnote{Available at http://www.jmmc.fr/aspro} and \texttt{SearchCal}\footnote{Available at http://www.jmmc.fr/searchcal} services co-developped by LAGRANGE and IPAG, and of CDS Astronomical Databases SIMBAD and VIZIER \footnote{Available at http://cdsweb.u-strasbg.fr/}.
\end{acknowledgements}

\bibliographystyle{aa}
\bibliography{SCrAN_interf}

\begin{appendix}

\section{Chromatic effects}
\label{App:chromaticity}
\begin{figure}[h]
    \centering
    \includegraphics[trim=5 0 40 30, clip, width=\linewidth]{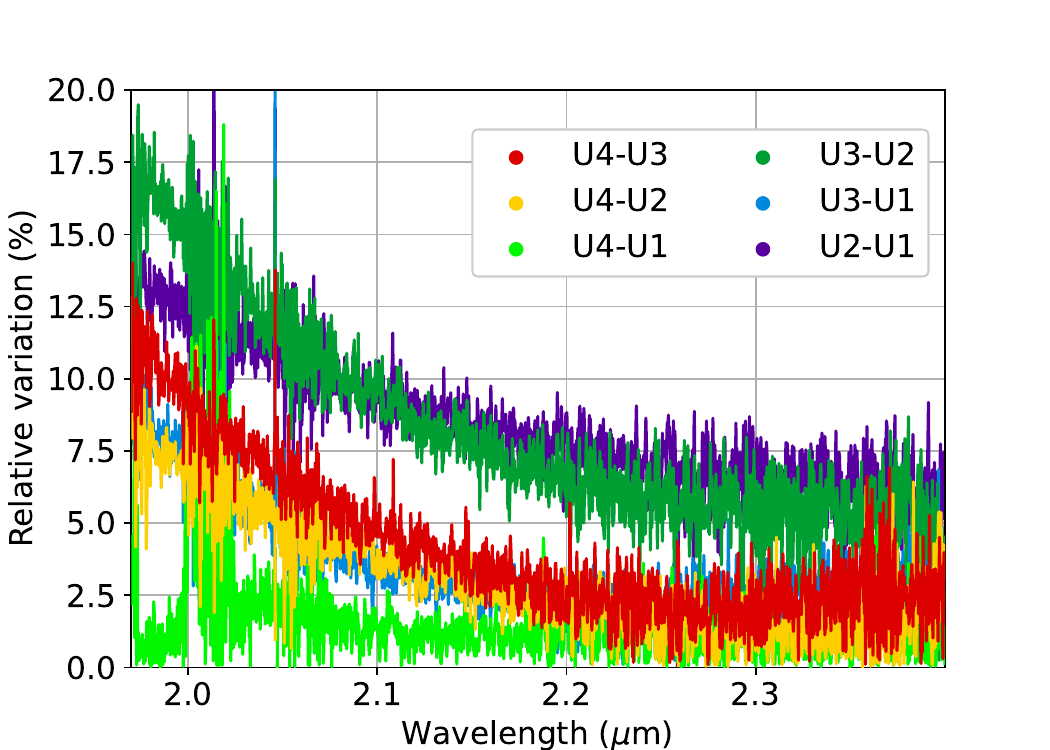}
    \caption{ Variation of squared visibility of the calibrator HD~186419 between our two epochs, as a function of wavelength, while colors code for the baseline.}
    \label{fig:Cal}
\end{figure}

The calibrator HD~186419 was observed on both nights at similar observing slots (i.e., at UT04:41 and UT04:45). As expected, the FT visibilities squared are identical for both nights. HD~186419 appears to be almost unresolved (with visibilities squared  $V^2$ around 0.90-0.98) and could be used to estimate the instrumental transfer function. Strikingly, while the FT tracked and locked the fringes during the whole observing sequences, the SC data appear different between the two nights (see the relative variation of visibility squared between the two epochs $V^2(20)/V^2(19)- 1$) in Fig.~\ref{fig:Cal}). Except for the longest baseline U1-U4, the visibility squared of August, 20 exhibits more pronounced curvatures with wavelength, and smaller values notably at the edges of the spectral range. The data also show an offset in visibility squared for the shortest baselines. Such behaviours could be explained by fringe jumps that affect more the wavelengths that are further away from the center of the spectral band because the spectral slope of the fringes is larger at the edges. The discrepancies appear also to be larger when UT2 and UT3 are involved, suggesting a worse performance of the adaptive optics and/or higher vibration effects on these telescopes specifically. Because such an effect might affect the absolute calibration of the visibility, we only used the data from the second night to constrain the dusty inner ring. This effect does not impact the differential observables across the \Brgam~line.


\section{Flux normalization with PMOIRED}
\label{App:PMOIRED}
\begin{figure}
    \centering
    \includegraphics[trim= 55 10 90 28, clip,width=\linewidth]{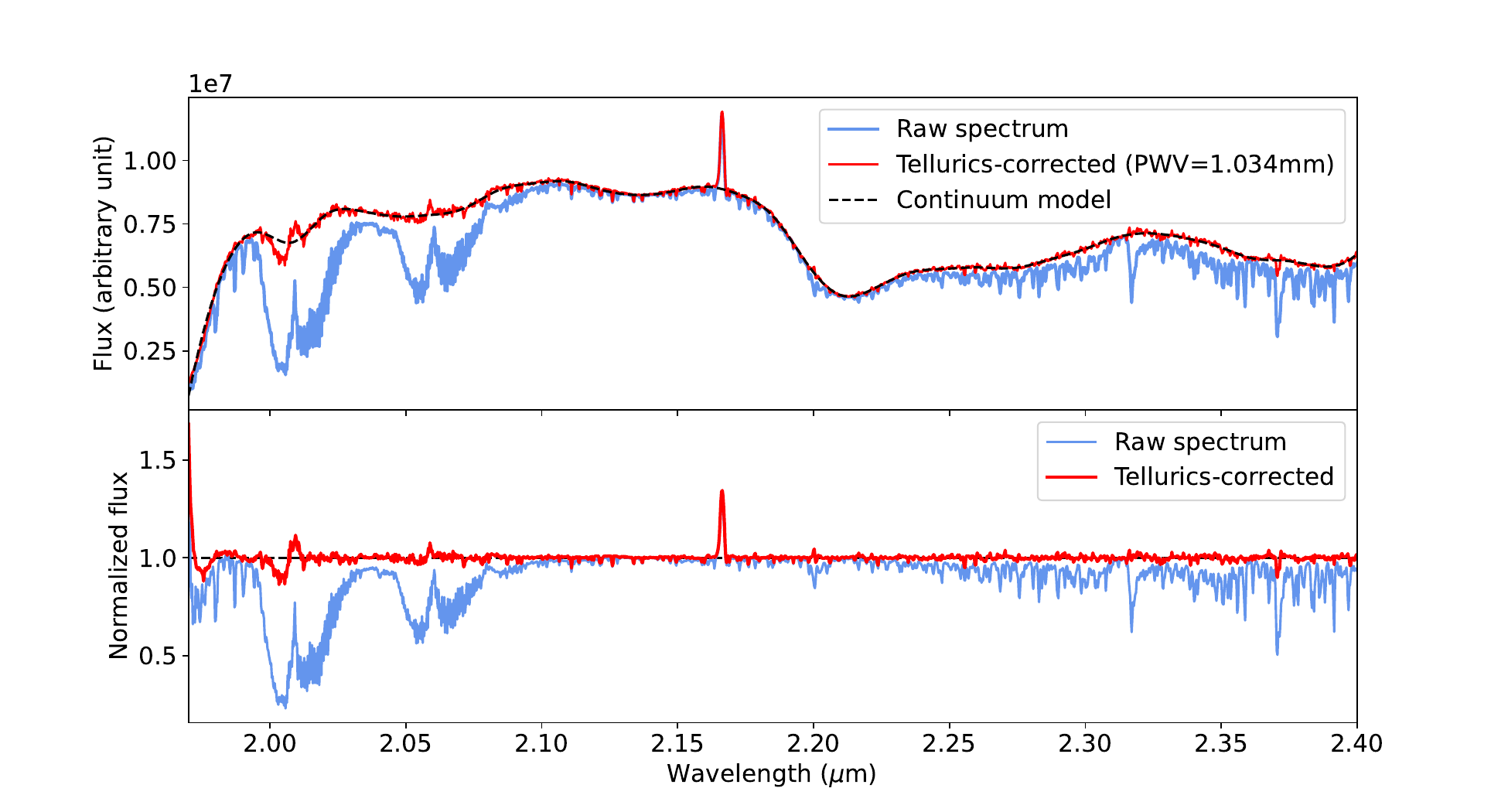}
    \caption{An example of S~CrA~N spectrum normalization with the PMOIRED tools (see text for details) on August, 20\Th. }
    \label{fig:tellurics}
\end{figure}

PMOIRED is a modeling tool particularly suited for GRAVITY data. Here, we highlight the use we made of the \texttt{tellcorr} Python3 library for flux normalization and wavelength calibration in the case of S~CrA~N. This process has been shown in one of the very handful examples of PMOIRED available online\footnote{The code in its original version, along with tutorials and full license can be found at https://github.com/amerand/PMOIRED}.\\
The library includes a 1D grid of atmospheric models. Its only parameter is the precipitable water vapour (PWV). The code performs a fit of the continuum spectrum over the whole K band. A telluric transmission function is inferred, along with a continuum model which is obtained thanks to a spline model of 35 nodes (for the high resolution mode of GRAVITY) fitted to the observed spectrum. Thanks to the telluric model obtained, a refined spectral dispersion of the instrument can be estimated, which we used for the present work. The normalized spectrum is finally obtained by dividing the observed spectrum by the continuum and the telluric model. This operation has been performed on all the spectra of each night. We illustrate the different aforementioned quantities in Fig.~\ref{fig:tellurics}, and give the derived PWV in Table~\ref{tab:PWV}.\\
\begin{table}[ht]
    \caption{Precipitable water vapor (PWV) of the atmospheric models that best reproduce the GRAVITY K band continuum for all spectra.}
    \centering
    \begin{tabular}{c|c|c|c|c}
        \hline
        \hline
        & \multicolumn{2}{c|}{August 19\Th} & \multicolumn{2}{c}{August 20\Th} \\
       \hline
       File nb & Obs. date & PWV & Obs. date & PWV\\
          & [MJD - 59810] & [mm] & [MJD-59810] & [mm] \\
       \hline
       1 & 0.1359 & 1.371& 1.1184 & 1.147\\
       2 & 0.1452 & 1.349& 1.1279 & 1.192\\
       3 & 0.1498 &1.349& 1.1325 & 1.173\\
       4 & 0.1549 &1.325& 1.1375& 1.154\\
       5 & 0.1594 &1.311& 1.1421& 1.156\\
       6 & 0.1689 &1.338& 1.1514 & 1.158\\
       7 & 0.1735 &1.338& 1.1560 & 1.175\\
       8 & 0.1784 &1.336& 1.1609 & 1.158\\
       9 & 0.1879 &1.331& 1.1655 & 1.128\\
       10 &- &-& 1.1748 & 1.092\\
       11 & -&-& 1.1794 & 1.067\\
       12 & -&-& 1.1843 & 1.034\\
       \hline\hline
    \end{tabular}
    \label{tab:PWV}
\end{table}

\section{A fully resolved halo in \Brgam}
\label{App:halo}
\begin{table*}
        \caption{Best-fit parameters for the disk-only model with 1$\sigma$ error bars. The Position Angle is from North to East.}
    \centering
    \begin{tabular}{c | c c c c | c c c c}
    \hline
    \hline
    $ $ & \multicolumn{4}{c|}{August 19} & \multicolumn{4}{c}{August 20} \\
    \hline
    Velocity & HWHM & $\alpha$ & $PA$ & $\chi_r^2$ & HWHM & $\alpha$ & $PA$ & $\chi_r^2$\\
    
    [km/s] & [mas] & [\ddeg] & [\ddeg] & [-] & [mas] & [\ddeg] & [\ddeg] & [-] \\
    \hline
    -190 & $0.89\pm0.05$ & $57.0\pm3.5$ & $154.9\pm3.6$ & 1.0 & $0.84\pm0.04$ & $56.1\pm3.2$ & $159.5\pm2.9$ & 1.6 \\
    -153 & $086\pm0.03$ & $59.3\pm2.2$ & $158.9\pm2.0$ & 0.97 & $0.85\pm0.03$ & $58.4\pm1.9$ & $160.5\pm1.6$ & 2.5 \\
    -117 & $0.85\pm0.02$ & $59.4\pm1.5$ & $160.0\pm1.2$ & 1.92 & $0.87\pm0.02$ & $60.4\pm1.1$ & $159.7\pm0.9$ & 4.1 \\
    -80 & $0.82\pm0.01$ & $58.8\pm1.0$ & $160.3\pm0.9$ & 2.0 & $0.87\pm0.01$ & $60.3\pm0.7$ & $158.5\pm0.5$ & 6.3 \\
    -44 & $0.79\pm0.01$ & $57.1\pm0.8$ & $158.8\pm0.7$ & 4.4 & $0.83\pm0.01$ & $59.4\pm0.5$ & $156.9\pm0.5$ & 10.5 \\
    -7 & $0.80\pm0.01$ & $56.6\pm0.7$ & $157.4\pm0.7$ & 5.1 & $0.83\pm0.01$ & $59.5\pm0.5$ & $156.5\pm0.4$ &  11.9 \\
    +30 & $0.80\pm0.01$ & $56.7\pm0.7$ & $156.5\pm0.6$ & 6.3 & $0.83\pm0.01$ & $59.3\pm0.5$ & $155.7\pm0.4$ & 12.5 \\
    +66 & $0.79\pm0.01$ & $56.3\pm0.8$ & $158.1\pm0.7$ & 6.0 & $0.82\pm0.01$ & $58.7\pm0.5$ & $155.7\pm0.5$ & 11.2 \\
    +103 & $0.79\pm0.01$ & $56.0\pm0.9$ & $159.2\pm0.8$ & 6.0 & $0.81\pm0.01$ & $57.6\pm0.6$ & $156.3\pm0.6$ & 10.1 \\
    +139 & $0.81\pm0.02$ & $56.4\pm1.2$ & $158.9\pm1.1$ & 2.5 & $0.80\pm0.01$ & $56.0\pm0.9$ & $157.2\pm0.8$ & 5.6 \\
    +176 & $0.77\pm0.04$ & $64.3\pm2.9$ & $161.0\pm2.4$ & 1.0 & $0.77\pm0.03$ & $63.2\pm2.6$ & $169.7\pm2.2$ & 1.9 \\
    \hline\hline
\end{tabular}
\label{tab:diskG}
\end{table*}

\begin{figure}[h]
    \centering
    \includegraphics[trim=380 0 0 0, clip,width=\linewidth]{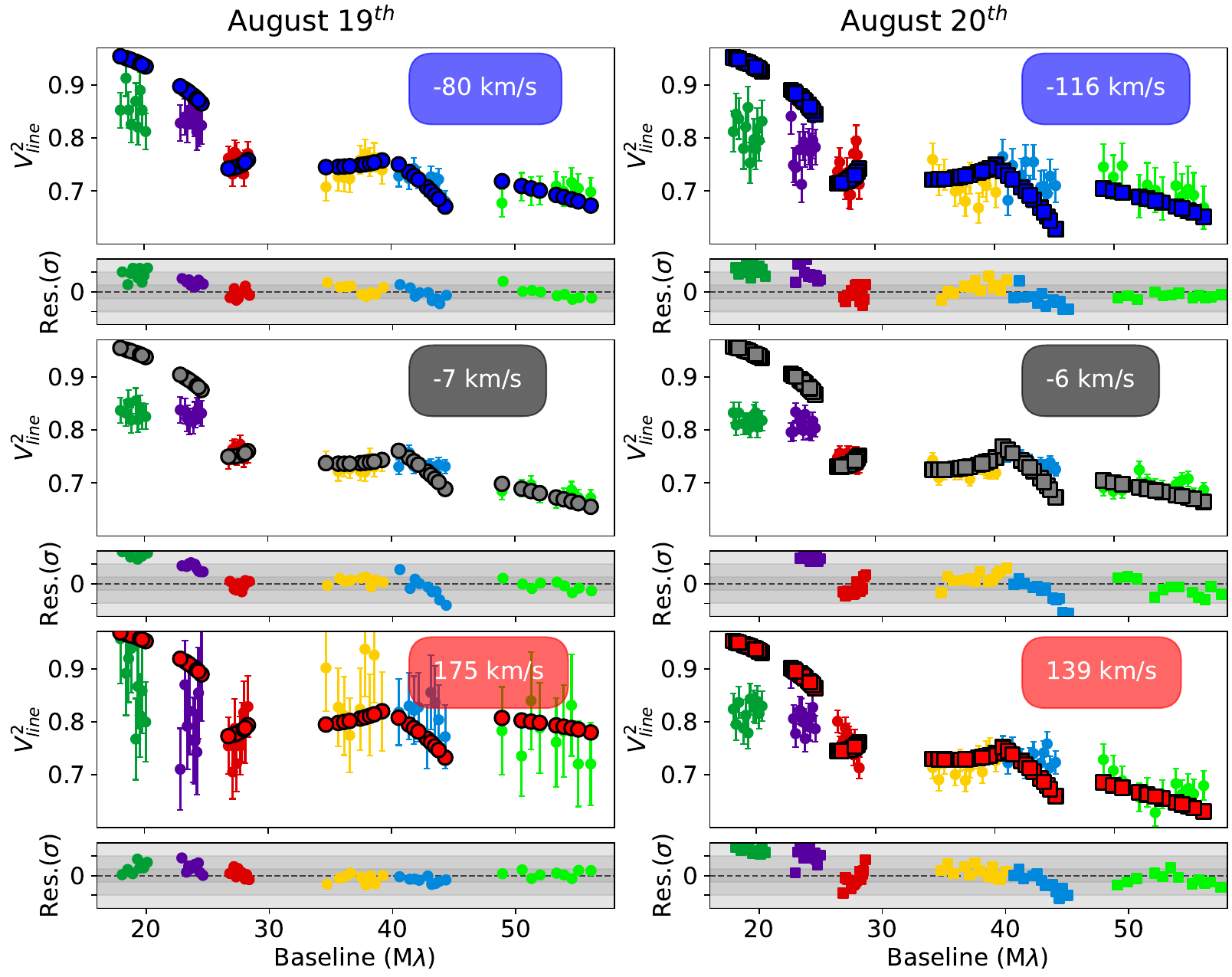}
    \caption{Same as Fig~\ref{fig:sizes-fit}, but with the halo's contribution forced to 0\%.}
    \label{fig:sizes_G}
\end{figure}

Prior to any fitting procedure, a cautious analysis of the pure-line visibility curves reveals that a simple Gaussian disk model is not suited to reproduce these data. The low frequency values are too low, and the high frequency values decrease too slowly with frequency. Both effects can be addressed with the addition of a fully resolved halo to the model. to illustrate the lacks of the simple Gaussian model, we show in Table~\ref{tab:diskG} and Fig.~\ref{fig:sizes_G} the results obtained when fitting the \Brgam~pure line visibilities with such a model. We can see that the residuals are much more dispersed (up to 5$\sigma$) than with a halo (see Fig.~\ref{fig:sizes-fit}). We also see systematic deviations of the model from the data. These deviations follow a clear trend for each baseline, regardless of the exact parameters obtained, suggesting that the inclined Gaussian disk model is not completely representative of the inner regions flux distribution. This claim is strengthened by the reduced $\chi^2$, which are a factor 2 to 12 higher than when considering a halo in the model (see Table~\ref{tab:disk}).
The implementation of a fully resolved halo requires to express the \Brgam~pure line visibilities $V_\text{line}$ as composite (just like for the continuum analysis) with two contributors : a disk and a halo. When considering $F_\text{D}$ and $F_\text{H}$, the disk and halo individual fluxes and $V_\text{D}$ and $V_\text{H}$ their respective visibility :
\begin{equation}
    V_\text{line} = \frac{F_\text{D}.V_\text{D}+F_\text{H}.V_\text{H}}{F_\text{D}+F_\text{H}}.
\end{equation}
By definition, $V_\text{H}=0$, while the visibility of the disk $V_\text{D}$ is given in, e.g., \citet{berger_2007} :
$$V_D = \exp{\left(-\frac{\left(2\pi\,\Theta\sqrt{u^2+v^2}\right)^2}{4\ln 2}\right)}$$
With $\Theta$ the Gaussian HWHM. And since $B/\lambda = \sqrt{u^2+v^2}$ :
\begin{equation*}
    V_\text{line} = \frac{F_\text{D}}{F_\text{D}+F_\text{H}}\exp{\left(-\frac{\pi^2B^2\,\Theta^2}{\lambda^2\ln 2}\right)}
\end{equation*}
We identify the contribution of the disk with respect to the total flux $C_\text{D}=\frac{F_\text{D}}{F_\text{D}+F_\text{H}}$ which can be expressed as a function of the contribution of the halo relative to the total flux $C_\text{H}~=~\frac{F_\text{H}}{F_\text{D}+F_\text{H}}~=~1-C_\text{D}$, for the sake of discussion. Hence the expression used in our modeling of the \Brgam~differential data (see main text) :
\begin{equation*}
    V_\text{line} = C_\text{D}\exp{\left(-\frac{\pi^2B^2\,\Theta^2}{\lambda^2\ln 2}\right)}, 
\end{equation*}
\begin{equation}
    V_\text{line} =  (1 - C_\text{H})\exp{\left(-\frac{\pi^2B^2\,\Theta^2}{\lambda^2\ln 2}\right)}.
\end{equation}
The main effect we observe from this modeling is a decrease by a factor $\sim 1.5$ of the size of the disk when putting $\sim 9$\% of the total flux into an unresolved halo.

\end{appendix}
\end{document}